\documentclass[prl,twocolumn]{revtex4}
\usepackage{bm}
\usepackage{graphicx}
\usepackage{amsmath}
\usepackage{eufrak}
\usepackage{color}
\usepackage{upgreek}
\usepackage{amssymb,amsmath}
\usepackage{amsfonts}
\usepackage{bm}
\usepackage{color}
\usepackage[unicode=true,colorlinks=true,citecolor=blue]{hyperref}
\usepackage{placeins}
\usepackage{lipsum}
\usepackage[normalem]{ulem}

\newcommand{\nix}[1]{}

\begin{document}

\title{Observation of anomalously strong penetration of terahertz electric field through terahertz-opaque gold films  into a GaAs/AlGaAs quantum well}

%%%\titlerunning{Observation of anomalously strong penetration of terahertz}

\author{S.~D.~Ganichev$^1$, S.~N.~Danilov$^1$, M.~Kronseder$^1$, D.~Schuh$^1$, I.~Gronwald$^1$, D.~Bougeard$^1$,  E.~L.~ Ivchenko$^2$, and A.~Ya.~Shul'man$^3$}

%%%\authorrunning{S.~D.~Ganichev, S.~N.~Danilov, M.~Kronseder \it{et al.}}
%%%\institute

\affiliation{$^1$Terahertz Center, University of Regensburg, 93040 Regensburg, Germany\\
$^2$ Ioffe Physico-Technical Institute, 194021 St. Petersburg, Russia\\
$^3$ Kotelnikov Institute of Radioengineering and Electronics RAS, 125009 Moscow, Russia}

%%%\maketitle

\begin{abstract}
We observe an anomalously high electric field of terahertz (THz) radiation acting on a two-dimensional  electron gas (2DEG) placed beneath a thin gold film, which, however, is supposed to be opaque at THz frequencies.  We show that the anomalously  strong penetration of  the THz electric field through a very high conductive gold film emerges if two conditions are  fulfilled simultaneously: (i) the film's thickness is less than the skin depth and (ii) the THz electric field is measured beneath the film at distances substantially smaller than the radiation wavelength. We demonstrate that under these conditions the  strength of the field acting on a 2DEG is almost the same as it would be in the absence of the gold film. The effect is detected for macroscopically homogeneous perforation-free gold films illuminated by THz-laser radiation with a spot smaller than the film area. This eliminates the near-field of the edge diffraction as a possible cause of the anomalous penetration. The microscopic origin of the effect remains unexplained in its details,  yet. The observed effect can be used for the development of THz devices based on two-dimensional materials requiring robust highly conducting top gates placed at less than nanometer distance from the electron gas location.
\end{abstract}

%\pacs{
 %73.21.Fg,
%Quantum wells
 %72.25.Fe Optical creation of spin polarized carriers
 %78.67.De,
%Quantum wells
 %73.63.Hs
%Quantum wells
%}
\maketitle

\section{Introduction}
\label{introduction}

It is hardly necessary to emphasize the importance of  thin metal films in modern science and technology.  Metal films with thickness ranging from  a few Angstroms  to tens of  nanometers has received great attention in recent years. They are used in solar cells, detectors, electronic semiconductor devices, magnetic memory devices,  optical coatings (such as anti-reflective coatings), chemical and biological sensors, plasmonic and terahertz devices etc. Optical properties of thin metal films with thickness substantially smaller than  the light wavelength $\lambda$ are usually characterized by light reflectance and transmittance measured at distances much larger than $\lambda$, in the so-called  far-field, for review see e.g.~\cite{Lynch1998}.

These properties are rather well researched. Recently an understanding of electromagnetic field characteristics at distances from metal films substantially smaller than the radiation wavelength has become increasingly significant. It is primarily caused by the development of near-field optics, see e.g.~\cite{Sipe1983,Paesler1996,Courjonbook2003,Ohtsu2010,Kawata2010,NovotnyHecht2012,Keller2012,Talebi2019}. Furthermore, metal films are often used in the fabrication of metallic gates, metamaterials, plasmonic and  near-field opto-electronic devices based on 2D materials (e.g. graphene, the most prominent example), topological insulators and low dimensional heterostructures. In these cases the electric field distribution in the subwavelength vicinity of  metal films plays a crucial role. It is known that the rough surface of a well-conducting metal film enhances the interaction of electromagnetic radiation with objects located near the film. The most famous example of this phenomenon in the visible band is the surface enhanced Raman scattering (SERS) of light by organic molecules on the  rough silver surface, when an increase takes place in the scattering cross section up to $10^5 - 10^6$ times presumably due to the to the plasma excitation in surface roughness  features \cite{10Moskovitz,11Kneipp}. In the terahertz band, a roughness of metal gate made of gold or aluminum film results in a nonlinear photoconductivity of a tunnel Schottky diode that corresponds to an increase in the local energy density of the electromagnetic field in some small region of the gate relative to the energy density in the incident radiation up to $10^3 - 10^5$ times due to the field enhancement in the near zone during radiation diffraction by surface irregularities ~\cite{Ganichev1996,13Shulman1999,ganichev_book,Olbrich2016}.
 
  For homogeneous highly conductive films, however, it is expected that the radiation is totally reflected and the electric field behind the film should decrease by orders of magnitudes. 

\begin{figure}[b]
    \centerline{\includegraphics[width=0.6\linewidth]{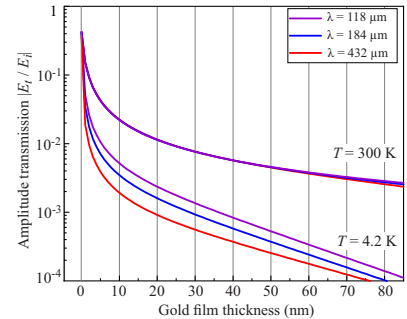}}
    \caption{ Ratio of transmitted and incident radiation electric field  as a function of the gold film thickness. Data are presented for three wavelengths and two temperatures.    }
 \label{1fresnel}
\end{figure}

\begin{figure}[htb]
    \includegraphics[width=\linewidth]{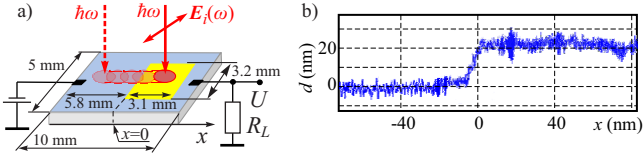}
    \caption{(a) Experimental setup. Vertical downward arrows and red spots  illustrate the polarized THz radiation beam scanned across the sample. The beam was scanned following the middle line along the long side of the sample. Step of 100, 200 or 250 $\mu$m were used. The geometrical sizes, being of the same order for all samples, and typical gold film profile are exemplary shown for sample \#B-20E. (b) Thickness profile $d(x)$ of 20~nm thick evaporated gold film measured by DektakXT Stylus Profiler (Bruker GmbH). Zero thickness $d$ corresponds to the surface of GaAs top cap layer.}
\label{2setup}
\end{figure}

Here we report on the observation of anomalously strong penetration of terahertz electric field through tens of nanometers thick homogeneous perforation-free gold films deposited on top of GaAs/AlGaAs heterostructures. Using the two-dimensional electron gas (2DEG) in GaAs quantum well as detector,  we show that the electric field of THz radiation acting on the 2DEG located beneath the highly conductive gold film is almost the same as it would be without the gold film, given that two conditions are fulfilled simultaneously: (i) the gold film has a thickness $d$ less than  skin depth $\delta_\mathrm{s}$ and (ii) the field is detected at a distance much less than the radiation wavelength $\lambda$. Taking into account the high  conductivity $\sigma$ of gold at terahertz frequencies~\cite{walther,grishkowsky,Lindquist,Lloyd}, this is a very surprising result.
 
Indeed, it follows from classical electrodynamics that, under condition $\omega/\sigma \ll  1$, where $\omega$ is the angular frequency and $\sigma$ is the conductivity, a metal film has a reflection coefficient $|r| \leq 1$ and a transmission coefficient $|t| \ll 1$, even at thicknesses much thinner than the skin depth $\delta_s = c /\sqrt{2 \pi \omega \sigma}$, where $c$ is the speed of light. The dependence of metal film penetrability on the thickness $d$ for the film placed on a dielectric substrate with the refractive index $n_b$ is given under conditions of the classical skin effect by 
 	\begin{equation}
 	\frac{E_t}{E_i} = \frac{ 2 \gamma t_0 {\rm e}^{- \beta d}}{ (1 - {\rm e}^{- 2 \beta d} ) + \gamma ( 1 + {\rm e}^{- 2 \beta d} ) }\:.
 	\end{equation}
 	Here $E_t$ and $E_i$ are complex amplitudes of the electric field in the transmitted and incident waves, ${t_0} = 2/(1 + n_b)$ is the amplitude transmission coefficient with no metal film, 
 	%
 	%%\begin{eqnarray}
 	%%&&\hspace{2.2 cm}\beta = \frac{ 1 - i}{\delta_s}\:, \nonumber\\
 	%%&&\gamma = (1 - {\rm i}) \frac{1 + n_b}{2} \sqrt{\frac{f}{\sigma}}  = (1 - {\rm i}) (1 + n_b) \pi \frac{\delta_s}{\lambda}\:, \nonumber
 	%%\end{eqnarray}
 	% 	\hspace{2.2 cm}
 	$$
 	\beta = \frac{ 1 - i}{\delta_s}\:, \gamma = (1 - {\rm i}) \frac{1 + n_b}{2} \sqrt{\frac{f}{\sigma}}  = (1 - {\rm i}) (1 + n_b) \pi \frac{\delta_s}{\lambda}\:,
 	$$
 	$f = \omega/2 \pi$ is the frequency, and $\lambda$ is the wavelength.
 
 Considering this equation for thin films characterized by $2 |\beta| d \ll 1$ we obtain in accordance with the impedance expression provided by Tinkham~\cite{Tinkham,Tu}
%%$$
\begin{equation}\label{transmit-thin-film}
t =\frac{E_t}{E_i} \cong  \frac{t_0}{1 + d\frac{\beta}{\gamma}} = \frac{t_0}{ 1 + \frac{\lambda}{(1+n_b) \pi \delta_s}\frac{d}{\delta_s} }
%%$$%\end{equation} 
%%\begin{equation}\label{transmit-thin-film}
= \frac{t_0}{ 1 + \frac{4 \pi d \sigma}{c(1 + n_b)} }\:.
\end{equation} 

From here follows  that in order to achieve a radiation transmission through the film close to unity, the  film thickness $d$ needs to be several orders of magnitude smaller than the skin depth $\delta_\mathrm{s}$, so that $d / \delta_\mathrm{s} \ll (1 + n_b) \pi \delta_\mathrm{s} / \lambda$. Accordingly, for the wavelength of about  $\lambda \approx 100~\mu$m (frequency $f \approx 3$~THz) an almost unchanged electric field amplitude is only expected for the gold film thickness $d < 1 \mbox{ {\AA}}$.  Consequently, the electric field at the second boundary as well as the amplitude of the transmitted electromagnetic wave at any real film thickness should be much less than the field amplitude in the incident wave. Noteworthy to note, that this abrupt decrease in the amplitude of the electric field inside the metal occurs immediately at the vacuum-metal interface, where the electric fields of the incident and reflected waves are almost mutually canceled  due to the high electrical conductivity of the metal. A further decrease in the field amplitude deep into the metal is characterized by the parameter $\delta_s$ and is little significant under our conditions $d <\delta_s$.  A. E.  Kaplan has made recently analogue conclusions by using also the impedance language~\cite{kaplan}.   Figure~\ref{1fresnel} shows the transmitted electric field amplitude, which has been calculated  after Fresnel formulae as a function of the gold film thickness for three different frequencies used in our experiments, for calculation details see Appendix~1.  

It demonstrates that for films with a thickness of 10~nm or more the field magnitude should decrease by at least two orders of magnitude. This has been observed in a number of experiments on far field transmission of THz radiation, see e.g. Ref.~\cite{walther}. It was also detected in our reference experiments performed under conditions $d_i \gg \lambda$ and/or $d > \delta_\mathrm{s}$, where $d_i$ is the distance from the film to the QW position at which the field is probed. The unchanged electric field amplitude, observed in our experiments for $d_i \ll \lambda$ and $d < \delta_\mathrm{s}$, disclose an obvious contradiction to fundamental textbook arguments given above. While the origin of this phenomenon unexplained in its microscopic details yet we believe that our experiments presented below may serve as basis for an understanding of a fundamental physical problem providing insight into a structure of electromagnetic fields at the metallic film - semiconductor interface.

The paper is organized as following. In Sec.~\ref{experimental} we briefly discuss the essential features of our research methodology and the properties of the gold-film/semiconductor structures under study. The results on THz fields penetrability are presented  in Sec.~\ref{results}.  The last section summarizes the work. We also supply our paper by Appendixes~1-3 presenting description of the equations used for calculations of curves in Figure~\ref{1fresnel}, experimental methods and additional details of growth and electric characteristics of the gold films.

\begin{figure}[tb]
    \centerline{\includegraphics[width=0.7\linewidth]{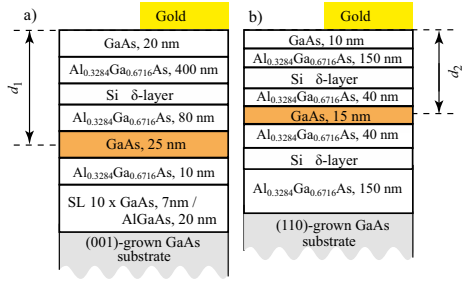}}
    \caption{Structure cross-section of wafers type \#A, panel (a), and \#B, panel (b), used  for fabrication of samples.  Single GaAs QW is grown at a distance $d_i$ from the top cap layer surface. Structures are grown on 350 $\mu$m thick GaAs substrate.}
    \label{3crosssectionQW}
\end{figure}

\begin{figure}[t]
    %%\vspace{-12mm}
    \includegraphics[width=\linewidth]{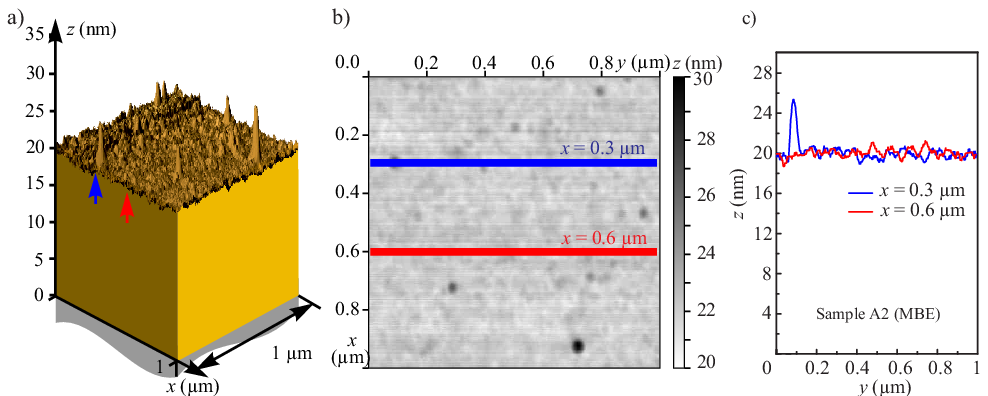}
    \includegraphics[width=\linewidth]{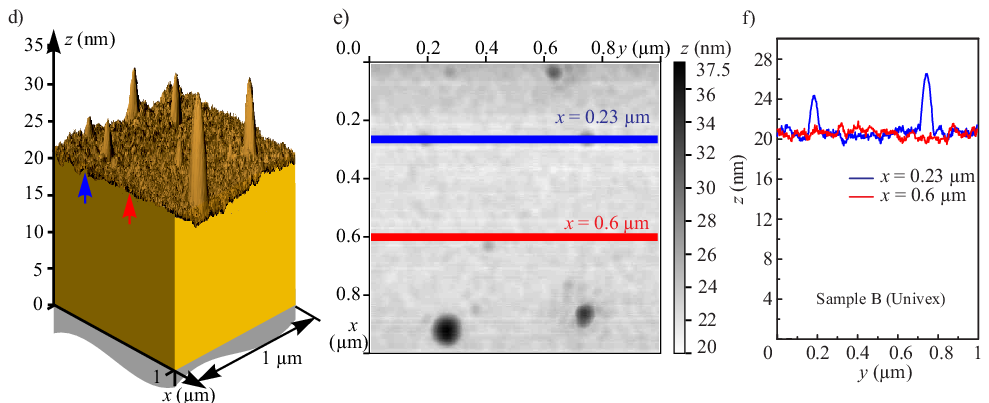}
    \includegraphics[width=\linewidth]{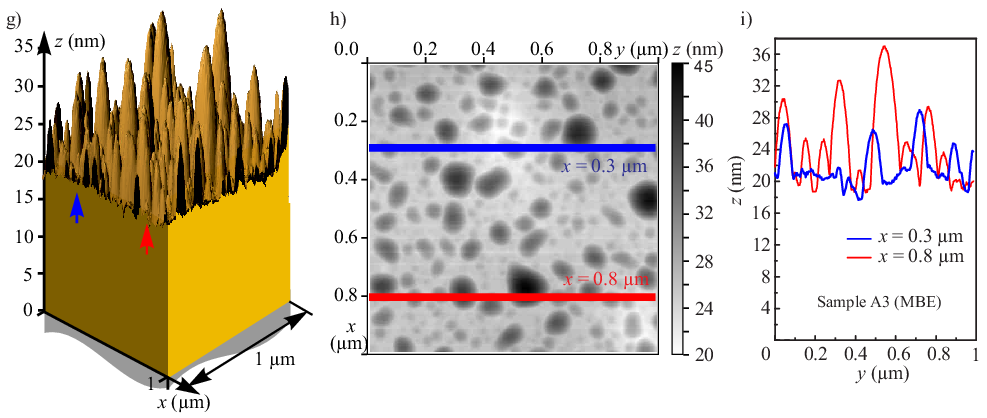}
    %%\vspace{10mm}
    \caption{ AFM profile of 20~nm gold layers measured for different samples. Images obtained on samples: (a)-(c) sample \#A1-20M(MBE grown gold); (d)-(f) -- sample \#B-20E (evaporated gold); and (g)-(i) -- sample \#A2-20M (MBE grown gold).}
    \label{4AFM}
\end{figure}

\begin{figure}[t]
     \centerline{\includegraphics[width=0.7\linewidth]{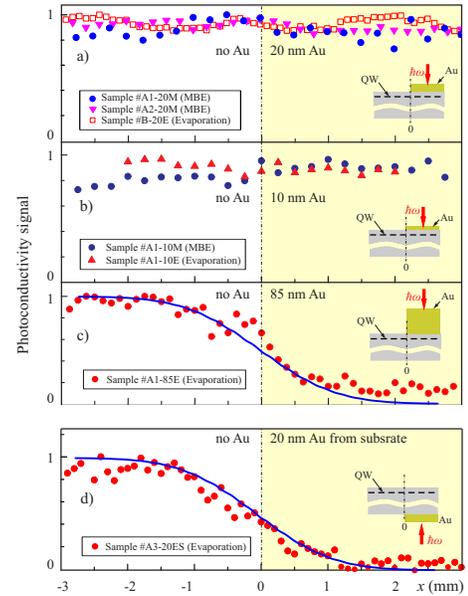}}
    \caption{ Coordinate dependence of the normalized photoconductivity signal in response to $cw$ radiation with $\lambda = 118~\mu$m obtained in samples with gold film of thickness; (a) 20~nm, (b) 10~nm, (c) 85~nm and (d)  20~nm. For three first panels, (a)-(c), gold films were fabricated on top of the heterostructure. In these cases normally incident THz laser radiation was applied from the top side. In the last panel, (d), gold film was deposited on the substrate side and normally incident radiation was applied from the substrate side.  Note that semiinsulating GaAs substrate are transparent for THz radiation. The data are normalized on the signal for the area without gold film. The laser beam was scanned along the axis $x$. The areas covered by gold are marked by yellow background. Solid lines in panels (c) and (d) show calculated integral intensity of a Gaussian beam scanned across the opaque film. Parameters of the Gaussian beam have been obtained applying piroelectric camera, see Appendix~2.} 
\label{5scansmain}
\end{figure}

\section{Experimental technique and samples}
\label{experimental}

Gold films are deposited on top of GaAs/AlGaAs heterostructures leaving either half of the sample uncovered or depositing gold only on a predefined rectangle,  see Figs.~\ref{2setup}(a) and~\ref{3crosssectionQW}. The anomalous penetrability of Au films have been observed by detecting the photoconductivity~\cite{ganichev_book} or the linear photogalvanic effect (LPGE)~\cite{ganichev_book,ivchenko05a} excited by THz radiation in a GaAs/AlGaAs quantum well (QW)  placed underneath the gold film at a distance $d_i$ much smaller than the radiation wavelength $\lambda$ in the GaAs ($d_{i} \ll \lambda$).  These effects yield photosignals with an amplitude proportional to the second power of the radiation electric field acting on the 2D electron gas in the QW, for details see Appendix~2. We compared the photosignals obtained for light illuminating either gold covered or gold uncovered areas of the sample. Taken the ratio of these signals, $\tilde{R}$, we obtain the  reduction of the radiation intensity caused by the gold film. The experimental geometry is shown in Fig.~\ref{2setup}(a). In particular, the beam having the spot size (diameter about 1.5~mm, see Appendix~2) smaller than the gold film area (about $3.5 \times 3.5$~mm$^2$) was scanned across the sample. Linearly polarized laser radiation was applied at normal incidence. Illuminating the sample changes the conductivity of the QW, which for biased sample resulted in a voltage drop across the load resistance $R_L$. In some samples it also yielded a photovoltage caused by the linear photogalvanic effect. The signal in response to the modulated radiation was registered by standard lock-in technique. Experiments were carried out applying continuous wave $cw$ THz molecular laser operating at wavelengths of 118, 184 or 432~$\mu$m~\cite{Kvon2012,olbrich2013}. Corresponding frequencies (photon energies) are $f= 2.54$~THz ($\hbar\omega =10.51$~meV), $f=1.62$~THz ($\hbar\omega =6.7$~meV) and $f=0.69$~THz ($\hbar\omega =2.87$~meV). Experiments were performed at room and liquid helium temperatures. More details are given in Appendix~2. 

\begin{figure}[t]
    \includegraphics[width=\linewidth]{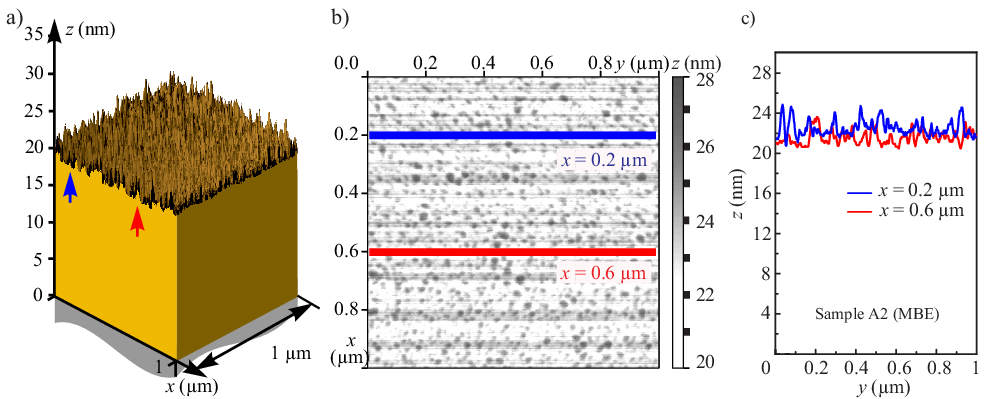}
    \includegraphics[width=\linewidth]{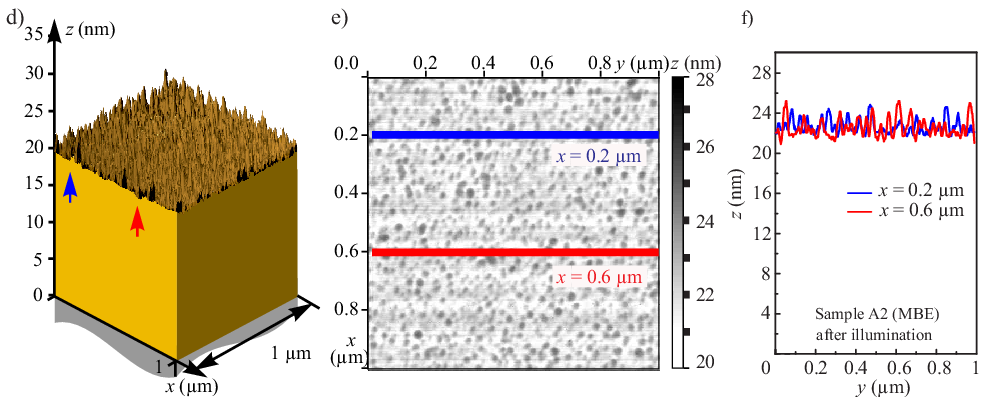}
    \caption{AFM profile of 20~nm MBE grown gold film (Sample \#A4-20M). Panels (a)-(c) -- before illumination and (d)-(f) -- after 35~min exposition to $cw$-laser radiation power$P \approx 115$~mW ($\lambda = 118\, \mu$m). }
    \label{6AFMillumin}
\end{figure}

Two groups of molecular beam epitaxially (MBE) grown GaAs/GaAlAs heterostructures with single GaAs quantum well were used. Figure~\ref{3crosssectionQW} shows a cross-section of the structure. Four wafers of the same heterostructure design,
%), used for samples \#A1-4,
% Wafers  \#A1, \#A2, and \#A3
% \#A1 (MC180130A)
% \#A2 (MC180801A)
% \#A3 (C030504A)
% \#A4 (MC181024A)
grown on (001)-oriented GaAs substrate and hosting a 25 nm thick QW, were used for fabrication of type \#A samples.
Wafer \#B, 
%(D060602B)
grown on a (110)-oriented GaAs substrate and hosting a 15~nm thick QW,  was used for fabrication of samples of type \#B. The latter structure, with a reduced point group symmetry compared to (001)-grown structures, allows us to measure besides the photoconductivity also the photogalvanic effect excited by normal incident THz radiation~\cite{footnote1} in unbiased samples, for details see Appendix~2. Substrates are made of 350~$\mu$m thick semiinsulating GaAs and are transparent for THz radiation. The density and mobility of electrons in 2DEG, obtained by magnetotransport experiments performed at $T=4.2$~K are  $n \approx 10^{11}$~cm$^{-2}$ and mobility $\mu \approx 10^6 \,{\rm cm^2/ V \cdot s}$, respectively.

Gold films with thicknesses 10, 20, and 85~nm were fabricated either by MBE directly connected to the GaAs-MBE or thermal vacuum evaporation. For the latter the samples were exposed to air before the Au deposition. The $3.5\times3.5$~mm$^2$ gold films covered about half of the sample surface with a rectangular shape with about 5~mm width and 10~mm length. The shape and position of the gold layer are shown in Fig.~\ref{2setup}(a).  For optoelectronic measurements, after the  gold film deposition, a pair of ohmic contacts was prepared in the middle of short edges, see Fig.~\ref{2setup}(a). 
 
MBE grown gold films were used for preparation of samples  made from wafers \#A1, \#A2 and \#A4. For that subsequently to the  semiconductor growth, the wafers were transferred in-situ from the III-V %%%MBE 
to the Metal/Oxide MBE where the Au layers were grown on parts of the 2''-wafer by a movable shadow mask between the sample and the Knudsen cell. Reflection high energy electron diffraction (RHEED) was used to check the crystalline quality of the layer, and to crosscheck the growth rate by RHEED oscillations. In the following these samples will be labeled as e.g. sample \#A1-20M, which shows that the sample is fabricated  from the wafer \#A1, has a 20~nm thick film deposited on top of the heterostructure cap layer (the number after hyphen) and it is grown by MBE (last letter). More details are given in Appendix~3A.

Thermal vacuum evaporation  was used for preparation of samples made from wafers \#A1, \#A3 and \#B.  Gold films were deposited on an unheated wafer using a Leybold Univex system. The thickness of the film was monitored by a quartz-microbalance. The samples were not annealed after deposition. A metallic shadow mask was used to cover only half of the sample. The gold structure size was chosen such that all sample edges are not covered to avoid short-cutting the QWs, see Fig.~\ref{2setup}(a). These samples are labeled in the same way as MBE samples, e.g. sample \#B-20E, but with last letter E in the notation, indicating that the film is made by vacuum evaporation. For some measurements aimed to study the gold penetrability for the distances of gold film to QW larger than the radiation wavelength ($d_i \gg \lambda$),  we prepared sample with 20~nm gold film deposited on the backside of the sample's substrate instead of the deposition onto the top layer. This fact is reflected in the sample label by an additional letter ''S'', i.e. sample \#A3-20ES.

Figure~\ref{4AFM} shows typical atomic force microscopy (AFM) images of the fabricated gold films. These data show that the thickness fluctuations in MBE grown films do not exceed 20\%  and in evaporated films is less than 25\%. Consequently, the AFM images ensure that there are no perforations of film at least up to the resolution limit of the AFM. Figures~\ref{4AFM}(g)-(i) show AFM image of sample \#A2-20M. While being grown  by MBE due to unknown reason this sample is characterized by large thickness fluctuations (at some points up to 80\%). We included this sample in the study in order to check how important the surface roughness is for the gold film penetrability.

  \begin{figure}[t]
     \centerline{\includegraphics[width=0.7\linewidth]{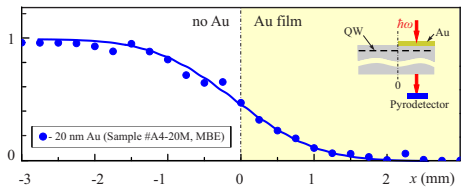}}
    \caption{Coordinate dependence of the  radiation transmission measured by pyroelectric photodetector placed directly after the  sample \#A4-20M with 20~nm (blue circles) gold film grown by MBE. The data are normalized on the signal for the area without gold film. Note that semiinsulating GaAs substrate are transparent for THz radiation. The data are obtained applying for $cw$  laser radiation with $\lambda 118\, \mu$m. Solid lines show calculated integral intensity of a Gaussian beam scanned across the opaque film. Parameters of the Gaussian beam have been obtained applying piroelectric camera, see Appendix~2.   }
    \label{7scanpyro}
\end{figure}

Besides the AFM measurements,  yielding gold film profiles on a micrometer scale,  we checked the film thickness over the whole sample length.  The results are shown in Fig.~\ref{2setup}(b). Here, the layer thickness was measured by DektakXT Stylus Profiler (Bruker GmbH). This is a  contacting method using low-force low-inertia scanning needle and has an accuracy of 4~{\AA}ngstr\"{o}m. The data show that the thickness variation is near $\pm 5$~nm and the border of the gold film is rather sharp ($\Delta x \simeq 10 \div 15 \mu$m).

\section{Results}
\label{results}

\subsection{Room temperature data}
\label{resultsRT}

Applying normally incident linearly polarized $cw$ radiation we detect an increasing resistivity of the QW's in all samples upon illumination (negative photoconductivity). The change of the QW's conductivity is caused by the electron gas heating due to Drude-like free carrier absorption and consequent change of the electron mobility (so-called hot electron bolometer or $\mu$-photoconductivity effect). The signal strength is proportional to the square of the amplitude of the electric field acting on the 2DEG. The negative sign of the  photoconductivity is due to the fact that at room temperature the mobility is governed by phonon scattering, which becomes more efficient upon heating. This phenomenon is well known and deeply studied for various bulk and low dimensional materials including GaAs QWs with characteristics similar to that used in our work, more details are given in Appendix~2. Thus, detailed discussion of the physical mechanisms of the photo-response is out of scope of the present paper aimed to anomalous penetrability observed in gold films.

\begin{figure}[t]
    \centerline{\includegraphics[width=0.8\linewidth]{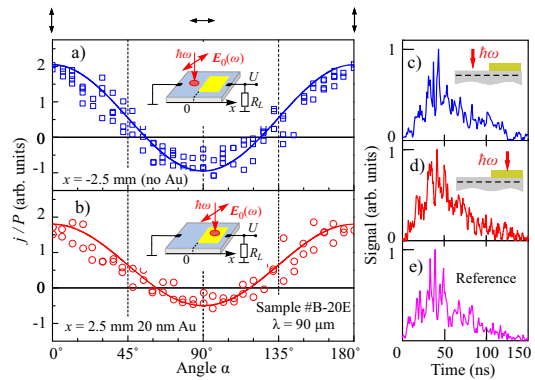}}
    \caption{(a) and (b): Photocurrent as a function of the azimuthal angle $\alpha$ measured in sample \#B-20E. Data are taken for irradiation of gold-film-free part (a) of the sample (blue points) and gold-film-covered part (b) (red points, film thickness 20~nm). Data are obtained with radiation of pulsed laser operating at wavelength $\lambda = 90\, \mu$m. The total pulse duration about 100~ns and peak pulse power $P\approx 800$~W. Solid lines are  fits by phenomenological equation~(\ref{alpha_normal}) of Appendix~2 describing LPGE and given  by $j/P =  \chi_- + \chi_+ \cos 2 \alpha$. Panel (c)-(e) show  pulse traces show photosignal for irradiation of gold-film-free-part (c), gold-film-covered part (d) of the sample and response of the reference photon drag detector (e).}
    \label{8lpge}
\end{figure}

Figures~\ref{5scansmain}(a) and (b) show  the main result of  the work obtained for 20 and 10~nm thick gold films deposited on the top surface. Moving the beam across the surface from the area without gold film to that covered by gold, surprisingly the photoconductive signal does not change much. It remains almost the same even if the spot is located fully within the gold film area! This anomalous penetrability was observed for films fabricated by both methods and having different surface roughness. We emphasize that the data are obtained for gold thicknesses at which according to Fresnel formulae the amplitude of the electric field penetrating through the film should be decreased by orders of magnitude, see Fig.~\ref{1fresnel} and Sec.~\ref{introduction}. A possible reason for this result could be film perforations or percolation effects. The latter was studied in detail by the transmission measurements of gold films of various thicknesses~\cite{walther}. It demonstrated that while radiation transmission is substantially enhanced for film with thickness up to 7~nm  the percolation effects for 20-nm thick film become absent. AFM images discussed above demonstrate that there are no perforations of film at least up to the resolution limit of the AFM. This makes near-field diffraction effect as a possible reason of the electric field enhancement very unlikely. We also note that the results for samples \#A1-20M with the most homogeneous gold film, see Fig.~\ref{4AFM}(a)-(c),  and \#A2-20M with the most rough film surface, see Fig.~\ref{4AFM}(g)-(i), are almost the same. Would the effect be caused by the film-surface roughness or near-field diffraction we would expect to detect difference in the electric field penetration in these samples with essentially different gold surface profile.

To  justify that our weak laser radiation does not modify the gold film  surface, e.g. drills holes in the films, we first took an AFM image at a marked position, than exposed the sample to the radiation for a long time (35 minutes), and finally again obtained the AFM image at previously marked position. For that we applied radiation with power levels even higher (115~mW) than that used for photoconductive measurements (40~mW). The results shown in Fig.~\ref{6AFMillumin} demonstrate  that the exposure with $cw$ laser radiation does not change the gold thickness and the surface roughness, excluding laser radiation induced gold ablation in our experiments. 

In the measurements discussed  above two conditions have been simultaneously fulfilled: (i) film thickness $d$ was less than the skin depth $\delta_\mathrm{s}$ (for used frequency about 65~nm at room temperature, see Appendix~3B), and (ii) the distance to QW-based detector, $d_i$, was much smaller than the wavelength $\lambda$  in GaAs.

Furthermore,  violation of any of these conditions results in a drastic reduction of the photoconductive signal demonstrating that the film becomes opaque. Figure~\ref{5scansmain}(c) shows the results obtained for a 85~nm gold film which is thicker than the skin depth. Scanning the beam from uncovered parts to gold-covered parts results in the substantial signal reduction. The data can be well fitted by a curve describing the intensity reduction expected for a Gaussian beam moving across the boundary of an opaque material, see curve in Fig.~\ref{5scansmain}(c). The Gaussian beam parameters used for the calculation were determined applying a pyroelectric camera (corresponding image and discussion are presented in the Appendix~2). 

\begin{figure}[t]
     \centerline{\includegraphics[width=0.7\linewidth]{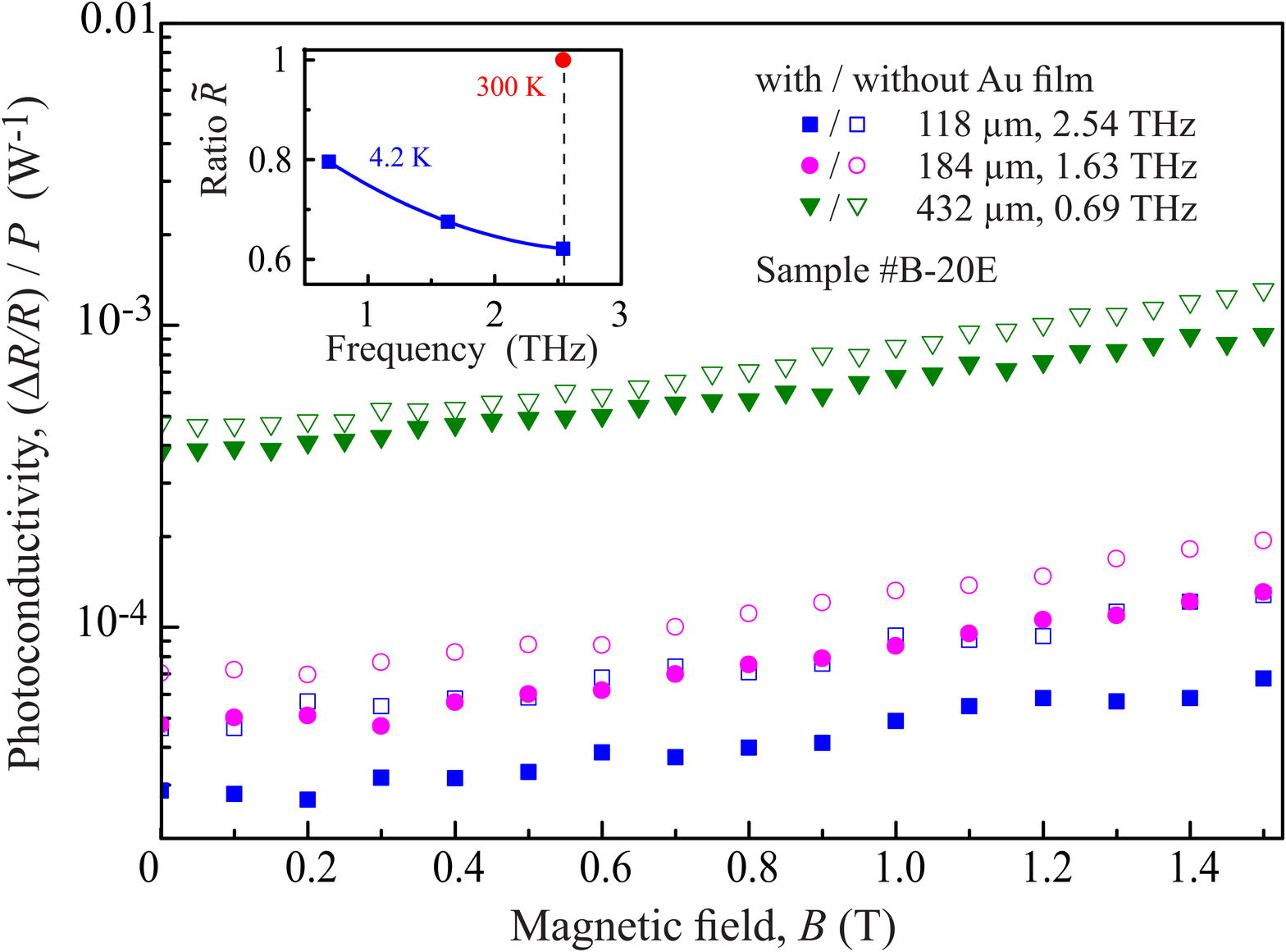}}
    \caption{Magnetic field dependence of the photoconductive signal  obtained for two spot positions with radiation focused on sample areas without and with 20~nm gold film. The data are obtained at liquid helium temperature. Inset shows frequency dependence  of the ratio $\tilde{R}$ of the signals obtained at zero magnetic field  for illumination of gold-film-covered and gold-film-free area of the sample.
    } \label{9lowT}
\end{figure}

Placing detector at a distance $d \geqq \lambda$ and, consequently, measuring the far-field transmission through films we confirm that, as expected from Fresnel formulae, our 20~nm films are opaque for THz radiation.  This has been verified by two methods. In the first one we deposited gold film on the bottom side of the heterostructure and illuminated the film from the substrate side, see inset in Fig.~\ref{5scansmain}(d).  In this case the distance between the gold film and QW detector (of the order of 350~$\mu$m) becomes larger than the radiation wavelength 118~$\mu$m.  Note that the latter is additionally reduced in GaAs by a refractive index of about 3.5. Figure~\ref{5scansmain}(d) shows the result for 20~nm thick gold film deposited on the backside of the substrate, sample \#A3-20ES. It demonstrates that the photoconductive signal  drastically reduces by shifting the beam to the gold film and the scan can be well fitted by intensity reduction at moving of the Gaussian beam across an opaque film.  In the second method we measured the the radiation transmission of a film for samples with 20~nm gold film placed on top of heterostructure. Now, however, instead of measuring photoconductive signal we measured a signal of a large area  pyroelectric detector placed behind the sample at several mm distance, i.e. much larger than  $\lambda$, see Fig.~\ref{7scanpyro}. The scans are shown in Fig.~\ref{7scanpyro} and demonstrate that the signal and, correpondently, radiation transmission become zero for a laser spot located within the gold-film area of samples.

In all measurements described above the gold film penetrability was studied by comparing  photoconductive signals obtained for radiation illuminating the gold-film-free and gold-film-coated areas of samples. It is noteworthy that the same result was achieved by using a completely different effect -- photogalvanic current excited in QWs by terahertz radiation. To obtain a significant signal, in this experiments we used radiation of pulsed THz laser yielding single 100~ns  pulses, see Appendix~2. The polarization dependencies of the photocurrent  generated by pulsed THz radiation in unbiased sample \#B for radiation spot positions at gold-film-free sample area and within the   20~nm gold film coated area are shown in Fig.~\ref{8lpge}(a) and (b), respectively. The data reveal that the magnitude of the signal and its polarization dependence are almost the same for both spot positions. This result is in accordance with the photoconductivity data obtained on the same sample, see Fig.~\ref{5scansmain}(a). Essentially new information obtained by LPGE  is that not only the radiation electric field magnitude acting on 2D electrons but also its polarization remains the same. Polarization dependence of the photocurrent,  described by  $j/P =  \chi_- + \chi_+ \cos 2 \alpha$ being in agreement with the phenomenological theory of the LPGE,   see Eq.~(\ref{alpha_normal}) in Appendix~2. The polarization dependence for the conditions relevant to our experiment, i.e., normal incidence excitation of (110)-oriented GaAs/AlGaAs QWs and the current measured along [110]-direction is discussed in Appendix~2, see Eq.~(\ref{alpha_normal}). The mechanisms of the LPGE in such QWs involve asymmetric scattering of photoexcited electrons, they are well known and will not be discussed here, for somewhat more details see Appendix~2.

To measure rather weak LPGE currents we applied pulsed laser radiation with possibly low  intensity allowing us  to avoid radiation induced gold film destruction. Using short radiation pulses additionally  allowed us to analyze the time constant of the photoresponse. Characteristic pulse traces,  shown in Fig.~\ref{8lpge}(c)-(e), demonstrate that the photoresponse for both spot positions reproduce short spikes caused by the spontaneous mode-locking (time constant less than 10~ns). Note that the same time constants were obtained for photoconductive signals measured  for illumination of areas without and with gold film (not shown). These short response times are in agreement with characteristics times of both of LPGE  and $\mu$-photoconductivity, see Appendix~2. The time scale as well as the coincidence of the time responses for the illumination of gold-film-free and and gold-film-coated sample parts additionally confirm that in both geometries  the origin of the photosignals is identical.

\subsection{Low temperature measurements}
\label{results4K}

Anomalous penetrability of terahertz electric fields through  20~nm gold films deposited on a GaAs quantum well structure was also observed at liquid helium temperatures. For these measurements the sample was placed in an optical cryostat with $z$-cut crystal quartz windows.  The photoconductive signal excited by normally incident radiation was measured as a function of an external magnetic field applied perpendicularly to the film. The data were obtained for three radiation wavelengths 118, 184 and 432~$~\mu$m corresponding to frequencies 2.54, 1.68 and 0.69 THz, respectively. We measured photoconductive signal for two sample positions at which radiation was focused on gold-film-covered or gold-film-free parts of the sample, see Fig.~\ref{9lowT}.  The values of the ratio $\tilde{R}$ were calculated as a ratio of these  signals. The inset in Fig.~\ref{9lowT} shows  the frequency dependence of the intensity reduction obtained  for zero magnetic field. In contrast to the room temperature data yielding for 20~nm thick film $\tilde{R} \thickapprox 1$, see the inset in Fig.~\ref{9lowT},  at low temperature $\tilde{R}$ varies from $\approx$0.6 to 0.8  by frequency decrease from 2.54 to 0.69~THz. Note that signals for both spot positions behave  equally upon magnetic field variation showing that the signals are caused by the same mechanism. The observed increase of the photoconductive signal with increase of magnetic field is attributed to magnetic field induced modification of scattering mechanisms and its discussion is beyond the scope of present paper. We also note that, besides the drastically larger electric field as compared to that expected from Fresnel formulas, the field reduction is characterized by  frequency dependence  opposite to that of the far-field gold film transmission, see Fig.~\ref{1fresnel}.

The observed temperature and frequency behaviour of the ratio $\tilde{R}$ qualitatively correspond  to that of the skin depth. Being aware that it can not be directly applied to our case since the electron mean free path is larger than the gold film thickness $d$ we, nevertheless, like to pay attention to these similarities. The skin depth $\delta_s$ can be re-written as
\begin{equation}
\delta_\mathrm{s} = \sqrt{\frac{\rho}{\pi f \mu_0}}.
\end{equation}

Due to the reduction of the film resistivity upon temperature decrease, see Appendix~3B, the skin depth at low temperatures is reduced as compared to that at  room temperature. Furthermore, the skin depth depends on  frequency as $\sqrt{1/f}$, i.e. becomes smaller for higher frequencies.  In 20~nm thick gold films at room temperature the skin depth, being of the order of 65~nm for $f=2.54$~THz (see Appendix~3B), is about three times larger than the film thickness. At 4.2~K, however, these values becomes comparable. Consequently, cooling the sample, results in partial violation of the inequality  $\delta_\mathrm{s} > d$, which  was shown above to be crucial for the anomalous penetrability.  Both tendencies, the decrease of the ratio $\tilde{R}$ with the temperature decrease at constant frequency or with frequency increase at constant temperature, are detected in our experiments, see inset in Fig.~\ref{9lowT}.  

\section{Summary}
\label{summary}
To summarize, we observe an anomalously strong penetration of THz radiation electric fields through almost homogeneous, perforation-free gold films deposited  on semiconductor heterostructures with a conductive quantum well placed beneath the film  at a distance substantially smaller than the radiation wavelength. We detected no field amplitude reduction as long as two conditions are simultaneously fulfilled: (i) the  thickness of a highly conductive film is less than the skin depth (about 65~nm at room temperature and for the frequencies used in the study) and (ii) the distance to the QW-based detector is much smaller than the wavelength. Our measurements demonstrate that a violation of any of these conditions results in a drastic reduction of the photosignal, showing that the film becomes opaque. In this case the result is described by Fresnel formula. So far we have no microscopic picture for this striking result. 
We note that, while we did our best to reduce the surface roughness, we cannot exclude completely inhomogeneities of the metal film, particularly on its internal surface, which would result in the effective conductivity to be less than expected by general frequency scaling law of DC bulk conductivity of the material. While focusing only on THz/gold/AlGaAs we bear in mind that the effect should be also present in other systems as far as as two above mentioned conditions are fulfilled simultaneously. At terahertz frequencies these conditions can be satisfied for various metals with high enough values of high-frequency conductivity, such as e.g, gold, silver, aluminum etc. For higher frequencies, i.e. in near infrared and visible range the conductivity decreases, and metals become less reflective, so that the field can penetrate through metal films of feasible thicknesses. As concerning the AlGaAs QWs used in our experiments, the only aim of it is a technical realization of local radiation electric field detection at nanometer scale distances from the metal film surface. Electrical photoresponse to terahertz radiation has been observed in a great number of low dimensional systems, different QWs and heterostructures, graphene, 2D materials etc., which offer a number of ways to realize such experiments. The observed effect seems attractive for the development of THz devices based on 2D materials, which  usually require robust top gates made of highly conductive metal films usually placed at less than one nanometer distance from the electron gas location.

\section{Acknowledgments} \label{acknow}

%%\begin{acknowledgments}
We thank  L. E. Golub for fruitful discussions.  The support from
the Deutsche Forschungsgemeinschaft (DFG, German Research
Foundation) - Project-ID 314695032 - SFB 1277,  the Volkswagen
Stiftung Program (97738) and the IRAP programme of the Foundation for Polish Science (grant MAB/2018/9, project CENTERA)  is gratefully acknowledged.
%%\end{acknowledgments}

%\FloatBarrier

\section*{Appendix 1: Transmission trough Au film}
\label{Appendix1Fresnel}

The data plotted in Fig.~\ref{1fresnel} are obtained from Fresnel formulae. The amplitude transmission coefficient through the system `vacuum (0) -- Au film (1) -- semi-infinite substrate (2)' is given by ~\cite{BornWolf}
\begin{equation}
t = {t_{10}t_{21} \text{e}^{i\phi} \over 1 - r_{01}r_{21}\text{e}^{2i\phi}}, \qquad \phi = 2\pi \sqrt{\varepsilon_1} {d\over \lambda}.
\end{equation}
Here $d$ and $\varepsilon_1$ are the thickness and the dielectric function of the Au film, $r_{ij}= -r_{ji}$ and $t_{ij}$ are the amplitude coefficients for reflection and transmission of the light falling from a half-infinite medium $i$ ($i = 0$ in vacuum and $i = 1$ in layer 1) on the half-infinite medium $j$, given by the Fresnel formulae for normal incidence:
\begin{equation}
t_{10} = {2 \over 1 + n_1}, \; t_{21} = {2n_{1}\over n_1 + n_{b}},  \;  r_{01} = {n_1-1\over n_1+1}, \;  r_{21} = {n_1-n_{b}\over n_1 +_{b}},
\end{equation}
%
%%\begin{eqnarray}
%%t_{10} = {2 \over 1 + n_1}, \qquad r_{01} = {n_1-1\over n_1+1}, \qquad \\ \nonumber
%%t_{21} = {2n_{1}\over n_1 + n_{b}},  \qquad r_{21} = {n_1-n_{b}\over n_1 + n_{b}},
%%\end{eqnarray}
%
where we introduced $n_1 = \sqrt{\varepsilon_1}$ and the refraction index of the substrate $n_{b} = \sqrt{\varepsilon_{b}}$.

For a metallic plate we have the dielectric function in the following form
\begin{equation}
\varepsilon_1 = 1 -{\omega_p^2\over \omega(\omega+i/\tau)}.
\end{equation}
For gold $\hbar \omega_p=8.5$~eV, and momentum relaxation time in metal $\tau=14$~fs~\cite{eps_gold} at room temperature and $\tau=238$~fs at liquid helium temperature~\cite{tau_4_2}. For the substrate we take the dielectric function of GaAs, $\varepsilon_{b}=13$.

\section*{Appendix 2: Radiation sources and experimental details}
\label{Appendix3experimental}

\begin{figure}[t]
	 \centerline{\includegraphics[width=0.8\linewidth]{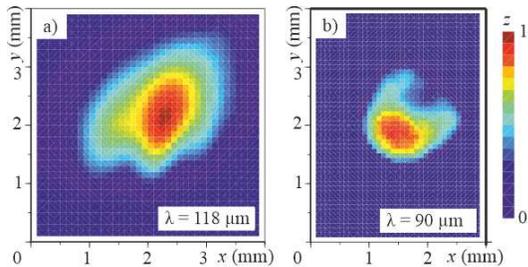}}
	\caption{Beam profiles of  (a) -- $cw$ laser operating at  $\lambda = 118\, \mu$m with the weighted mean beam diameter of 1.68~mm; (b) -- pulsed  laser operating at $\lambda = 90\, \mu$m with weighted mean beam diameter of 1.29~mm. }
	\label{14beamprof}
\end{figure}

Experiments were carried out applying continuous wave $cw$ THz molecular laser~\cite{Kvon2012,olbrich2013} operating at wavelengths of 118, 184 or 432~$\mu$m. Corresponding frequencies (photon energies) are $f= 2.54$~THz ($\hbar\omega =10.51$ meV) and $f=1.62$~THz ($\hbar\omega =6.7$ meV) and $f=0.69$~THz ($\hbar\omega =2.87$ meV). These single laser lines were obtained with methanol and difluormethane as active media. The radiation power on samples was about 40~mW. The radiation was modulated by a chopper at a frequency of $f \approx 130$~Hz. The beam cross-section had the Gaussian shape which was monitored by a pyroelectric camera~\cite{ganichev1999}. The spot sizes at the full width at half maximum (FWHM) are:  1.5~mm for $\lambda = 118~\mu$m, $2$~mm for $\lambda = 184~\mu$m and 3~mm for $\lambda = 432~\mu$m. Image for radiation with  $\lambda = 118~\mu$m is shown in Fig.~\ref{14beamprof}(a). They are smaller than the size of the $5\times5$~mm$^2$ square shaped gold film. To control the incidence power of the laser terahertz radiation a pyroelectric detector was used.  In an additional experiment, aimed to measure kinetic of the photoconductive signal,  we also used  a pulsed $NH_3$ laser~\cite{ganichev1998,ganichev2002}, operating at $\lambda=$ 90.5~$\upmu$m ($f= 3.31$~THz, $\hbar\omega =13.70$ meV). The laser was optically pumped by transversely excited atmospheric pressure (TEA) $\mathrm{CO}_2$ laser~\cite{ganichev2003,ganichev2007}. Single pulses with a  duration of the order of 100~ns and peak power about several kW were used. To control the incidence power of the lasers the photon drag detector~\cite{ganichev1985} was used. An example of the beam cross-section obtained with the pyroelectric camera is shown in Fig.~\ref{14beamprof}(b). The spot has a smaller size than gold films. The photoconductivity set-up differs from that used for $cw$ by the value of the load resistance $R_L=50\; \Omega$ only. The photovoltage across the load resistance was detected by digital oscilloscope. In  experiments on LPGE we controllable rotated  the polarization plane of linearly polarized radiation by the azimuth $\alpha$.  For that we used crystal quartz lambda-half plate  Note that for $\alpha =0$ radiation is  polarized along $y$-direction.

To conclude on the  gold film penetrability we compared amplitudes of photoconductive or photovoltage signals  obtained for illumination of gold film with that caused by illuminating of the gold-film-free area, see Fig.~\ref{2setup}(a).

Irradiation of QWs results in a  free carrier (Drude-like) absorption of THz radiation resulting in an electron gas heating. An increase  of the electron temperature $T_e$ changes the electron mobility $\mu$ which can be detected as a photoconductive signal. If heating is weak, the radiation-induced variation of the conductivity $\Delta\sigma$ can be well approximated by the simple expression~\cite{ganichev_book}
\begin{eqnarray}\label{e66}
\frac{\Delta\sigma}{\sigma_d}= \frac{1}{\mu}\frac{\partial \mu}{\partial T_e}\bigg\vert_{T_e=T}\Delta T,
\end{eqnarray}
where $T$ is  the lattice temperatures,  $\Delta T  = T_e -T$ and $\sigma_d$ is the conductivity without irradiation. We readily see that the sign of the photoconductive signal is determined by that of the derivative $\partial \mu / \partial T_e$. In the case of predominant  scattering by phonons, which is relevant to our experiment,  the sign of the photoconductivity is negative~\cite{Seeger1977}. In a biased structure photoexcited change of sample resistance results in a  $dc$ current proportional to the second power of the radiation electric field $E$ at frequency $\omega$ and the static field $\bm E(\omega = 0)$~\cite{ganichev_book}:
\begin{equation}
\label{photocond}
j_\alpha = \sigma_{\alpha \beta\gamma\delta }^{(3)} E_\beta(\omega) E_\gamma^*(\omega) E_\delta(\omega = 0),
\end{equation}
with conductivity $\sigma_{\alpha \beta\gamma\delta }^{(3)} \equiv \sigma^{(3)}_{\alpha \beta\gamma\delta}(\omega,-\omega,0)$. The $dc$ current caused by the increase of QW resistance is detected in a circuit sketched in Fig.~\ref{2setup}(a) as a voltage drop across load resistance. The response time of $\mu$-photoconductivity is determined   by fast times of  free carrier energy relaxation, which depends on temperature lie  in the ps to ns range.  Thus  excitation by 100~ns pulses with ns-time scale spikes caused by spontaneous mode-locking results in a fast photoconductive response, which repeats the  temporal structure of the radiation pulse. More details on the mechanisms of THz photoconductivity can be found in, e.g., Ref.~\cite{ganichev_book}.

Due to symmetry arguments, photoconductivity is the only optoelectronic effect, which can be excited in our large homogeneous (001)-oriented QWs (wafers of type \#A) excited by THz radiation at normal incidence~\cite{ganichev_book,ivchenko05a}. In (110)-grown QWs (wafers of type \#B) the symmetry of QW is reduced to C$_{2v}$ point group, which makes possible generation of photogalvanic effects in unbiased samples even at normal incidence~\cite{Shalygin2006,ganichev2014}. This linear photogalvanic effect has been used in our work to provide additional evidence for the anomalous gold film penetrability applying an independent method. The LPGE arises in homogeneous samples under spatially     homogeneous optical excitation. It is due to the  ``built-in''  symmetry properties of the media interacting with the radiation     field and is caused by an asymmetry in {\bm$k$}-space of the    carrier photoexcitation and of the momentum relaxation due to scattering of free carriers on, e.g., phonons in noncentrosymmetric crystals, for reviews see~\cite{ganichev_book,ivchenko05a,sturmanBOOK,ivchenkopikus}. The linear photogalvanic current density $\bm j$  is phenomenologically described by the following expression~\cite{ivchenko05a}
    \begin{equation}
    \label{Ch7LPGEphenomenolog} j_{\lambda} =\sum_{\mu, \nu}  \chi_{\lambda \mu \nu} \frac{1}{2} (E_\mu E^*_\nu + E_\nu E^*_\mu
    )
    \end{equation}
linking the $dc$ current  to the symmetrized product $\{E_\mu E^*_\nu\}$ by a third rank tensor  $\chi_{\lambda\mu\nu}$, which is symmetric in the last two indices.  The index $\lambda$ enumerates two in-plane coordinates of QW $x$ and $y$, while $\mu$ and $\nu$ run over all three Cartesian  coordinates.  Therefore, $\chi_{\lambda\mu\nu}$ is     isomorphic to the piezoelectric tensor and may have nonzero     components in media lacking a center of symmetry. The linear photogalvanic effect represents a microscopic ratchet. The periodically alternating electric field superimposes  a     directed motion on the thermal velocity distribution of carriers    in spite of the fact that the oscillating field does not exert a    net force on the carriers or induce a potential gradient. The   directed motion is due to nonsymmetric random relaxation and scattering in the potential of a noncentrosymmetric medium~\cite{Haenggi2009,weber2008,Ivchenko2011}. The polarization dependence of the LPGE current density $j_x$ for normal incidence by linearly polarized radiation follows from Eqs.~\eqref{Ch7LPGEphenomenolog}  and is given by~\cite{Shalygin2006,Diehl2007,belkov2003}
\begin{equation}\label{alpha_normal}
j_{x} = \left( \chi_{+} + \chi_{-}  \cos{2 \alpha} \right)  P\:,
\end{equation}
where $\chi_{\pm} = (\chi_{xyy} \pm \chi_{xxx})/2$, $x$ is parallel  to the  mirror reflection plane of QW, radiation power $P \propto E^2$, and $E$ is the amplitude of the radiation electric field exciting 2DEG. The polarization dependence  describe well our experimental data with fitting parameters $\chi_{\pm}$, see solid lines in Fig.~\ref{8lpge}(a) and (b) for the illumination of gold-film-free and gold-film-coveed parts of the sample~\#B. The response time of the LPGE is determined by the fast momentum relaxation time. Thus also for the LPGE current  excitation by 100~ns pulses with ns-time scale spikes caused by spontaneous mode-locking results in a fast response, which repeats the  temporal structure of the radiation pulse, see Fig.~\ref{8lpge}. Microscopic details of the LPGE in GaAs QWs excited by THz radiation are well researched and described in e.g.~\cite{ganichev_book,ivchenko05a}.

\section*{Appendix 3: Gold films fabrication and characteristics}
\label{Appendix2gold}

\subsection*{A. MBE gold films growth}
\label{Appendix2goldfabrication}

\begin{figure}[t]
    %%\vspace{-12mm}
    \includegraphics[width=\linewidth]{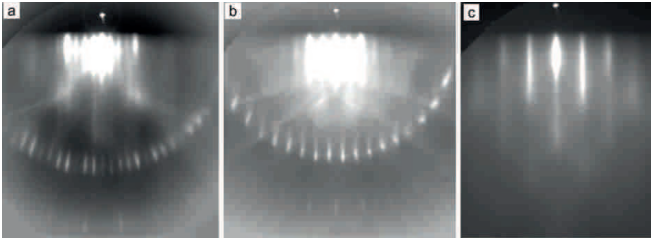}
    %%\vspace{10mm}
    \caption{RHEED images for (a)  GaAs(001) along the (100), (b) along the (011) direction and (c) for 20~nm Au grown on GaAs(001)}
    \label{10RHEEDKronseder1}
\end{figure}

\begin{table*}[t]
    \centering
    %\begin{tabular}{|c| c| c| c| c| c| c| c|}
    \begin{tabular*}{\textwidth}{l@{\extracolsep{\fill}}c c c c c c c c}
        \hline
        Sample & & Current & $U_{12,34}$  &  $U_{23,41}$ &  $R_{12,34}$ &  $R_{23,41}$ &  $\rho$ \\ 
         & & ($\mu$A) & ($\mu$V) &  ($\mu$V) &  $(\Omega)$ &  $ (\Omega)$ &   $(\Omega \cdot m)$ \\ \hline
        \#A3-20M & MBE & 10 & 5.07 & 5.03 & 0.507 & 0.503 & $4.6 \cdot 10^{-8}$ \\
        \#A4-20M & MBE & 10 & 5.52 & 4.94 & 0.552 & 0.494 & $4.8 \cdot 10^{-8}$\\
        \#B-20E & Evaporated & 10 & 6.52 & 6.14 & 0.652 & 0.614 & $5.8 \cdot 10^{-8}$ \\
        \hline
    \end{tabular*}
    \caption{Specific resistivity $\rho$ of 20~nm Au films measured by van-der-Pauw method. Current, voltages and resistances are values with that the surface resistance was obtained. Measurements applying lower current yielded the same result.        }
   \label{tab2}
\end{table*}

Samples \#A1, \#A2, and \#A4 were made out of metal/semiconductor  hybrid films grown completely in a MBE cluster. This means that after the semiconductor growth, the wafers were transferred in-situ from a semiconductor chamber to dedicated metal chamber, where the Au layers were grown on parts of the 2''-wafer using mobile shadow mask. The base pressure of the metal chamber is around $5 \times 10^{-11}$~mbar. Reflection high energy electron diffraction (RHEED) was used to check the crystalline quality of the layer, and to crosscheck the growth rate by RHEED oscillations, see Figure~\ref{10RHEEDKronseder1}. For gold,  oscillations with 40~s/ML were obtained. The thickness of the grown layers was additionally controlled by a calibrated quartz microbalance. In Figure~\ref{10RHEEDKronseder1}(c) the thin streaky RHEED pattern, here for 20nm Au layer grown on wafer \#A4,
%(MC181024A),
indicates the high crystalline quality of the Au-layer grown at $200^\circ$C substrate temperature. 

$X$-ray photoemission spectroscopy  (XPS) from a 40~nm Au films revealed that the topmost layer consists of Au, O, C, Ga and As. O and C is known to be present at all surface exposed to air. Ga and As is also known to be present in the topmost layer~\cite{1,3} after growth of Au directly on GaAs. In the initial growth stage, the Au atoms seem to get completely incorporated into the very first Ga and As layers, where a phase separation into Au-Ga phases may take place, triggered by the impinging Au atoms. After growth of a few nm, a pure Au layer is formed between the thin  surface layer containing Au, Ga and As and the underlying semiconductor heterostructure~\cite{1,3,2}. By Ar-ion bombardment at 1000 eV for 15 min. which removes approximately 3-6 nm from the top, the XP-spectra shows a pure Au layer without any contaminants, which means that the Ga and As contamination is not present throughout the whole Au layer. Hence, this result confirms a floating Ga-Au-As layer on top of a pure Au film.

\begin{figure}[t]
     \centerline{\includegraphics[width=0.6\linewidth]{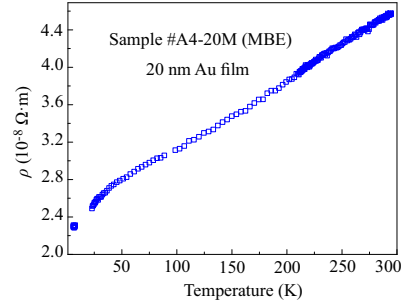}}
    \caption{Temperature dependence of the specific resistivity $\rho$ measured on a MBE grown 20~nm gold film. }
    \label{11rhoT}
\end{figure}

\subsection*{B. Electric characteristic of gold films}
\label{Appendix2rho}

Electric characteristics of gold  were obtained by measuring resistivity of MBE grown 20~nm film. The resistivity was determined  applying van-der-Pauw method~\cite{vanderPauw}. For that we used samples with square shaped 2x2 mm$^2$ gold films of 20~nm thickness. Four contacts to the film were fabricated with by silver-conducting varnish. Using the fact that in our samples the distance between contacts is substantially larger than the film thickness we calculated the specific resistivity after~\cite{vanderPauw}
\begin{equation}\label{vdPauw}
\rho = \frac{\pi d}{ln 2}\frac{R_{12,34}+R_{23,41}}{2}F
\end{equation}
here $d$ is the film thickness (20~nm),  $R_{12,34} = V_{34} / I_{12}$ and $R_{23,41} = V_{41} / I_{23}$  are resistances, $V_{34}, V_{41}, I_{12}$ and $ I_{23}$ are voltages and currents measured from the corresponding contacts, and factor $F = 1$, because in our measurements $R_{12,34} \approx R_{23,41}$.

Room temperature resistivity of various films obtained by this method are  given in Tab.~\ref{tab2} together with the values of applied currents, measured signals and calculated resistances for different pairs of contacts. The obtained values, being of the order of $4.6-5.8 \cdot 10^{-8} \,\Omega \cdot \rm{m}$, are close to that obtained in Ref.~\cite{grishkowsky} for 85~nm thick film ($8.7 \cdot 10^{-8} \,\Omega \cdot \rm{m}$). Both values are several times larger  than the value of the bulk gold resistivity $2.25 \cdot 10^{-8} \,\Omega \cdot \rm{m}$, see, e.g. \cite{Palik_HandbOptCnstSolids1998}. Note that the increase of specific resistivity with the film thickness decrease is well known, see e.g.~\cite{walther}. Using the specific resistivity from  Tab.~\ref{tab2} we obtain for room temperature and $f=2.54$~THz the skin depth  $\delta_\mathrm{s}\approx 65$~nm.The decrease of temperature results in a decrease of the specific resistivity and, consequently, of the skin depth. 

Temperature dependence of the gold film specific resistivity is shown in Fig.~\ref{11rhoT}. The data were obtained applying 10~$\mu$A  $(ac)$ current with  frequency 12~Hz. The corresponding voltage drop was measured by lock-in amplifier SR830.


\begin{thebibliography}{99}
\bibitem{footnote1} Note that in (001)-grown QWs structures \#A1, \#A2, \#A3 and
\#A4 photogalvanic effects at normal incidence are  forbidden by
symmetry arguments.
\end{thebibliography}
\end{document}